# Taming the Memory Beast: Strategies for Reliable ML Training on Kubernetes

Jaideep Ray
jaray@acm.org
Dec 16, 2024

## Abstract

Kubernetes offers a powerful orchestration platform for machine learning (ML) training, but memory management can be challenging due to specialized needs and resource constraints. This paper outlines how Kubernetes handles memory requests, limits, Quality of Service classes, and eviction policies for ML workloads, with special focus on GPU memory and ephemeral storage. Common pitfalls such as overcommitment, memory leaks, and ephemeral volume exhaustion are examined. We then provide best practices for stable, scalable memory utilization to help ML practitioners prevent out-of-memory events and ensure high-performance ML training pipelines.

## Introduction

Kubernetes (k8s) is a powerful container orchestration platform that automates the deployment, scaling, and management of containerized applications. Due to scalability, efficient resource management (GPUs, CPUs and memory) and portability (cloud, on-premise, hybrid), it has become a platform of choice for machine learning training workloads. One of the core functions of k8s is resource management, ensuring that containers receive the necessary resources like GPU, CPU and memory to function correctly.

- **CPUs:** When a container tries to use more CPU than allocated, Kubernetes throttles its usage, effectively slowing it down. This flexibility makes the CPU a compressible resource.
- **GPUs:** GPUs, on the other hand, are assigned to pods as whole units. A pod either gets the entire GPU or none of it. There's no concept of "throttling" GPU usage like with CPUs. This makes GPUs an inelastic resource in the context of Kubernetes.
- **Memory:** Training workloads usually RAM (main memory) and disk storage are considered incompressible resources in Kubernetes. This means that, unlike CPU resources, they cannot be throttled or shared flexibly among pods. When a pod requests a certain amount of memory or disk space, Kubernetes must guarantee that the full amount is available to the pod. Once allocated, it cannot be easily reclaimed, making it crucial to set requests and appropriate limits and monitor usage carefully. If a pod runs out of either main memory or disk, it can lead to out of memory (OOM) crashes, data corruption, or unpredictable behavior. If multiple pods sharing a node use more memory or disk available on the node, it will result in the k8s controller evicting pod(s) to restore stability. Since, Machine learning training jobs are increasingly using a large number of features and data, they can consume significant memory resources (RAM and disk). In this article we dive deep into areas which trigger the memory related crashes and related problems such as Out-of-memory killed error (OOMKilled).

## 2.1. Memory requests & limits

In Kubernetes, memory allocation is primarily governed by two key parameters: requests and limits. These parameters, defined at the container level, provide a mechanism for managing memory consumption and ensuring efficient resource utilization within a cluster.

- **Requests:** This parameter specifies the minimum amount of memory that a container requires to run. When scheduling pods, Kubernetes considers the requests of all containers within a pod to determine if a node has sufficient resources to accommodate the pod [1]. Kubernetes also considers the state of nodes, including their available CPU and RAM, when making scheduling decisions. It aims to place pods on nodes with sufficient resources to meet their resource requests [2]. If a node's available memory falls short of the combined requests of a pod's containers, the pod will not be scheduled on that node [3]. Kubernetes uses requests to determine if a node has enough resources to schedule a pod. If no requests are set, Kubernetes will assign requests equal to limits [4].

- **Limits:** This parameter sets the maximum amount of memory that a container can consume. Limits are enforced at runtime by the kubelet, preventing containers from exceeding their allocated memory [5].

At runtime, Kubernetes enforces limits (through Linux kernel cgroups) to ensure that containers do not exceed their allocated resources. Incorrect sizing of requests will result in scheduling delays and limits will result in OOMKiller being triggered resulting in pod termination.

A recommended approach is to set the memory request to the median observed memory usage and the memory limit to the peak usage and a small buffer. This strategy helps ensure that the training job has enough memory resources while preventing excessive overallocation.

To prevent disruption to neighboring jobs, it's important to align a training job's memory requests and usage with its GPU reservation. For instance, a job utilizing half of a node's available GPUs should ideally consume less than half of the node's RAM and disk resources. Note this is just a best practice as training jobs can be compute bound, memory bound or I/O bound depending on workload and data type.

## 2.2. Eviction due to memory pressure

Engineering teams put significant effort into maximizing the utilization of GPU nodes due to their high cost. It's common for multiple training workloads to run on the same GPU node.
In situations where multiple pods are running on the same node and collectively exceed the available memory, resource contention can occur. This can lead to performance bottlenecks, delays in training processes, and potential pod evictions.

Kubernetes allows for flexible resource utilization by enabling memory overcommitment. This means that the total memory requests of all pods on a node can be greater than the node's actual memory capacity. This flexibility stems from the fact that containers often don't use their full requested memory all the time [7]. It's important to remember that while overcommitment allows for this flexibility, Kubernetes prioritizes pods based on their Quality of Service (QoS) classes, especially when memory resources are scarce [8].

To manage overcommitment and prioritize resource allocation, Kubernetes uses QoS classes. These classes categorize pods based on their resource requirements and provide a mechanism for determining which pods to evict when memory resources become scarce [9, 10]. The three QoS classes [11] and eviction policies are as follows:

- **Guaranteed:** Pods in this class have their resource requests equal to their limits for all containers. These pods are guaranteed to receive the requested resources and are the least likely to be evicted during resource contention [10].
- **Burstable:** Pods in this class have at least one container with resource requests that are not equal to their limits. These pods are allowed to use more resources than their requests if available but may be evicted if they exceed their requests and the node experiences memory pressure.
- **BestEffort:** Pods in this class have no resource requests or limits specified for any of their containers. These pods have the lowest priority and are the first to be evicted when memory resources are oversubscribed [12].

Due to the varying memory demands of ML training jobs, it's advisable to avoid the BestEffort Quality of Service (QoS) class.

In addition to node-pressure eviction, Kubernetes also employs preemption eviction. This occurs when a higher-priority pod cannot be scheduled due to insufficient resources. Kubernetes will then evict lower-priority pods to make room for the new pod [13, 14].

For example, imagine a node with 16GB of memory with two pods running(pod A has higher priority than pod B):

- **Pod A:** Request: 4GB, Limit: 8GB, QoS Class: Burstable
- **Pod B:** Request: 8GB, Limit: 12GB, QoS Class: Burstable

If Pod A is currently using only 2GB of memory, the remaining 2GB of its requested memory can be used by Pod B, even though Pod B has a higher request. However, if Pod A suddenly needs its full 4GB, Kubernetes will evict Pod B if there's not enough free memory on the node.

If multiple workloads would be sharing the same GPU node, it is advisable to keep similar workloads co-located. This prevents noisy-neighbor issues.

## 2.3. Unexpected Spikes in Memory Usage

Training workloads can experience unexpected spikes in memory usage due to various operations such as dataset loading, feature pre-processing, shuffling batches in memory, caching, checkpoint writing etc.
Unexpected spikes in memory usage can result in job termination through OOMKill or pod eviction due to increased node pressure.

Analyze and track the memory usage of training jobs, especially during periods of high memory consumption. Memory spikes can be reduced by implementing careful programming and out-of-memory events can be reduced by establishing appropriate memory limits.

## 2.4. Memory leaks

The training job might have a bug that causes it to continuously allocate memory without releasing it. This can happen in both GPU memory or main memory. As memory usage reaches or crosses the limits, OOMKiller will eventually terminate the job.

Common sources of memory leaks in training jobs include: failing to release tensors, unnecessarily retaining or rebuilding computation graphs, and recreating large model components or layers in a loop without proper cleanup.

ML training often runs for many epochs or iterations. If a small leak happens on every iteration and is never reclaimed, the memory footprint will grow cumulatively. For stable training, it is essential to proactively detect memory leaks. Unexpected spikes and memory leaks can be understood by using tools like Tensorboard profiler.

## 2.5. GPU CUDA OOM

High Bandwidth Memory (HBM) is the fast, on-package memory used by modern GPUs. Running out of HBM will cause training job to fail. On HBM OOM, ML framework (TensorFlow, PyTorch, etc.) throw an explicit error message like the following:
- CUDA out of memory [15]
- Allocation failed
- cudaErrorMemoryAllocation

Mitigating HBM OOMs is about ensuring your workload stays within the physical constraints of the GPU's memory. Lowering batch sizes, data and model sharding are some techniques to reduce memory usage. It is also preferable to choose larger GPU HBM types for large model training. Larger GPU HBM types are preferred for training larger models.

## 2.6. Ephemeral storage OOM

Ephemeral storage is a type of volatile storage that is tightly coupled with a pod and does not persist once that pod is stopped.
Examples range from the local SSDs attached to compute instances to Kubernetes ephemeral volumes attached to pods. Because ephemeral storage capacity is typically limited and intended only for temporary data (for example, caches or logs), it can be exhausted [16] if an application writes more data than the ephemeral volume can hold. When that happens, you effectively get an OOM event for ephemeral storage, which can lead to pod eviction in containerized environments.

Ephemeral storage exhaustion in ML training jobs can be caused by various factors such as loading large model files or writing training logs and checkpoint files. To avoid this issue, identify the root cause of the storage issue and leverage persistent volumes instead.

## 3. Best Practices for Memory Management for training jobs

We have seen several causes of memory related pod crash and eviction. To ensure efficient and reliable memory management in Kubernetes, consider the following best practices:

- **Tune memory requests and limits appropriately:** Always define memory requests and limits for your pods to provide Kubernetes with the necessary information for scheduling and resource allocation[14].
- **Tune QoS depending on job criticality**: For critical jobs, use Guaranteed Qos as much as possible. For standard workloads, set requests to the median memory usage and limits as the peak memory usage.
- **Monitor Memory Usage:** Regularly monitor the memory (main memory, GPU HBM, Ephemeral storage, disk) usage of your pods to identify potential overcommitment or resource contention issues. Identify memory leaks or unexpected spikes early.
- **Prefer stable memory usage**: Stable memory usage is preferred for ML training jobs because it helps avoid runtime errors, ensures reproducibility, and simplifies resource planning. Unstable memory usage can lead to unexpected performance degradation, operating system issues, and out-of-memory (OOM) errors. This will interrupt training in the middle of a run and waste expensive compute.

## 4. Conclusion

Kubernetes provides a flexible and robust framework for managing memory resources in a containerized environment.

By carefully defining memory requests and limits, utilizing QoS classes effectively, monitoring memory usage and implementing jobs to have stable memory usage you can prevent out-of-memory issues, ensuring the stability and performance of your model training.

## 5. References

1. Efficient Resource Allocation Strategies in Kubernetes - LoadForgeGuides
https://loadforge.com/guides/efficient-resource-allocation-in-kubernetes
2. How Kubernetes CPU and RAM resources allocation works with Qovery? - Deployment
https://discuss.qovery.com/t/how-kubernetes-cpu-and-ram-resources-allocation-works-with-qovery/1354
3. Resource Management for Pods and Containers - Kubernetes
https://kubernetes.io/docs/concepts/configuration/manage-resources-containers/